\pdfoutput=1

\documentclass[11pt,twoside,a4paper,cmspaper,final,collab]{cms-tdr}
\def\svnVersion{360576}\def\cmsCernNoTag{CERN-EP/2016-007}\def\cmsCernDate{\today}\def\cmsMessage{Published in Physics Letters B as \href{http://dx.doi.org/10.1016/j.physletb.2016.06.080}{\doi{10.1016/j.physletb.2016.06.080.}}}

\begin{document}\cmsNoteHeader{SMP-14-019}

\hyphenation{had-ron-i-za-tion}
\hyphenation{cal-or-i-me-ter}
\hyphenation{de-vices}
\RCS$Revision: 344843 $
\RCS$HeadURL: svn+ssh://svn.cern.ch/reps/tdr2/papers/SMP-14-019/trunk/SMP-14-019.tex $
\RCS$Id: SMP-14-019.tex 344843 2016-05-30 17:01:04Z alverson $
\newlength\cmsFigWidth
\ifthenelse{\boolean{cms@external}}{\setlength\cmsFigWidth{0.85\columnwidth}}{\setlength\cmsFigWidth{0.4\textwidth}}
\ifthenelse{\boolean{cms@external}}{\providecommand{\cmsLeft}{top\xspace}}{\providecommand{\cmsLeft}{left\xspace}}
\ifthenelse{\boolean{cms@external}}{\providecommand{\cmsRight}{bottom\xspace}}{\providecommand{\cmsRight}{right\xspace}}

\ifthenelse{\boolean{cms@external}}{
\renewcommand{\Em}{\ensuremath{E\hspace{-0.55em}/\kern0.1em}\xspace}
\renewcommand{\ETslash}{\ensuremath{\Em_{\mathrm{T}}}\xspace}
\providecommand{\Emsub}{\ensuremath{E\hspace{-0.38em}/\kern0.05em}\xspace}
}
{
\renewcommand{\Em}{\ensuremath{E\hspace{-0.55em}/\kern-0.08em}\xspace}
\renewcommand{\ETslash}{\ensuremath{\Em_{\mathrm{T}}}\xspace}
\providecommand{\Emsub}{\ensuremath{E\hspace{-0.38em}/\kern-0.05em}\xspace}
}

\cmsNoteHeader{SMP-14-019}
\title{Measurement of the $\cPZ\gamma\to\cPgn\cPagn\gamma$ production cross section in $\Pp\Pp$ collisions at $\sqrt{s}=8\TeV$ and limits on anomalous $\cPZ\cPZ\gamma$ and $\cPZ\gamma\gamma$ trilinear gauge boson couplings}

\date{\today}

\abstract{
 An inclusive measurement of the $\cPZ\gamma\to\cPgn\cPagn\gamma$ production cross section in $\Pp\Pp$ collisions at $\sqrt{s} = 8\TeV$ is presented, using data corresponding to an integrated luminosity of 19.6\fbinv collected with the CMS detector at the LHC. This measurement is based on the observation of events with large missing energy and with a single photon with transverse momentum above 145\GeV and absolute pseudorapidity in the range $\abs{\eta} < 1.44$. The measured $\cPZ\gamma\to\cPgn\cPagn\gamma$ production cross section, $52.7 \pm 2.1\stat \pm 6.4\syst \pm 1.4\lum\unit{fb}$, agrees well with the standard model prediction of $50.0^{+ 2.4}_{- 2.2}\unit{fb}$. A study of the photon transverse momentum spectrum yields the most stringent limits to date on the anomalous $\PZ\PZ\gamma$ and $\PZ\gamma\gamma$ trilinear gauge boson couplings.
}

\hypersetup{%
pdfauthor={CMS Collaboration},%
pdftitle={Measurement of the production cross section for Z gamma to nu nu gamma in pp collisions at sqrt(s) = 8 TeV and limits on Z Z gamma and Z gamma gamma triple gauge boson couplings},%
pdfsubject={CMS},%
pdfkeywords={photon, MET}}

\maketitle
\section{Introduction}

The study of the production of boson pairs provides an important test
of the electroweak sector of the standard model (SM), since this production is a
consequence of the non-Abelian nature of the underlying $SU(2)\times U(1)$ symmetry.
Trilinear gauge boson vertices are a consequence of this symmetry, and the values of the self-couplings
are fixed in the SM. Any measured deviation would be an indication of physics beyond the standard model
at that vertex. For production of a $\PZ$ boson and a photon, these
couplings are zero in the SM. New symmetries or new particles that
only become relevant at higher energies could result in a cross
section that differs from the SM prediction~\cite{vvreview, baur_original}, particularly for
final-state bosons with high transverse momentum.

In this letter a measurement is presented of the production of a $\PZ$ boson, which decays into a pair of neutrinos, and a photon in proton-proton collisions, at a centre-of-mass energy of $\sqrt{s} = 8\TeV$, using data collected by the CMS experiment corresponding to
an integrated luminosity of 19.6\fbinv. This result extends previous measurements at the LHC~\cite{atgc7tev,ATLAS7TeV,ll8TeV}. We describe
a measurement of the production cross section as well as the extraction
of limits on anomalous $\cPZ \mathrm{V} \gamma$ couplings, where $\mathrm{V}=\cPZ,\gamma$. In this search for anomalous trilinear gauge couplings (aTGCs), the final-state boson transverse momentum is used as a sensitive observable.

\begin{figure}[htb]
\begin{center}
\includegraphics[width=0.45\textwidth]{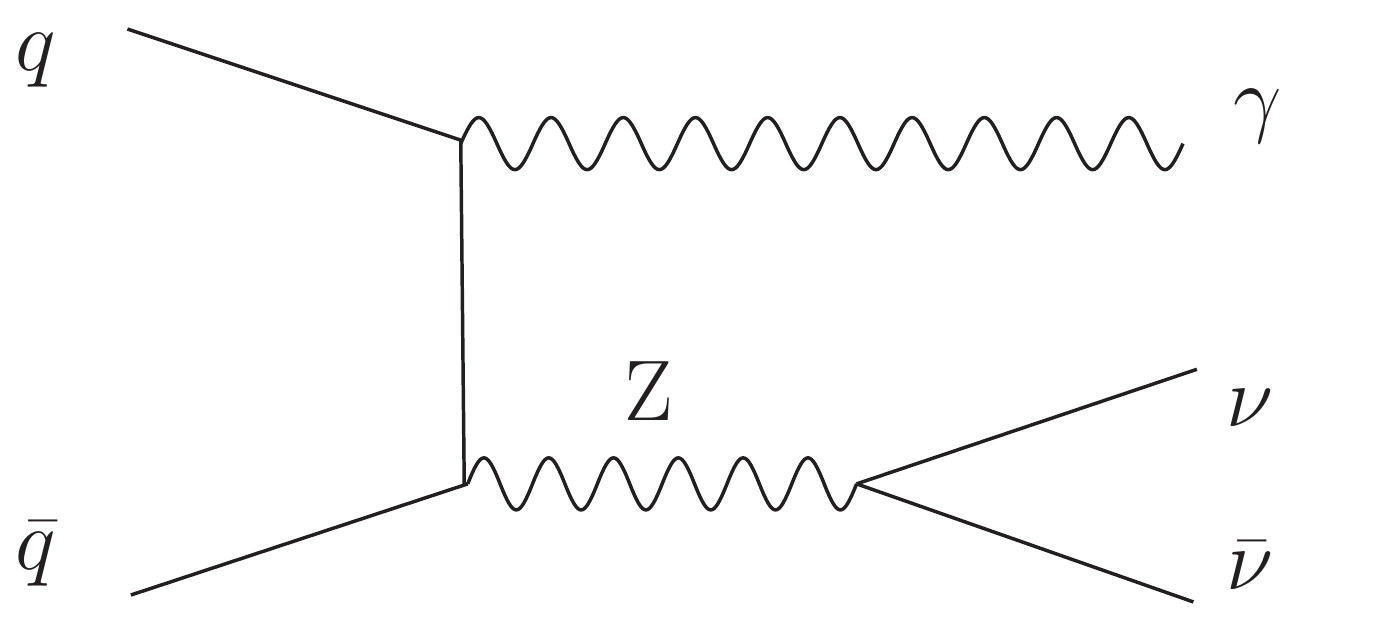}
\includegraphics[width=0.45\textwidth]{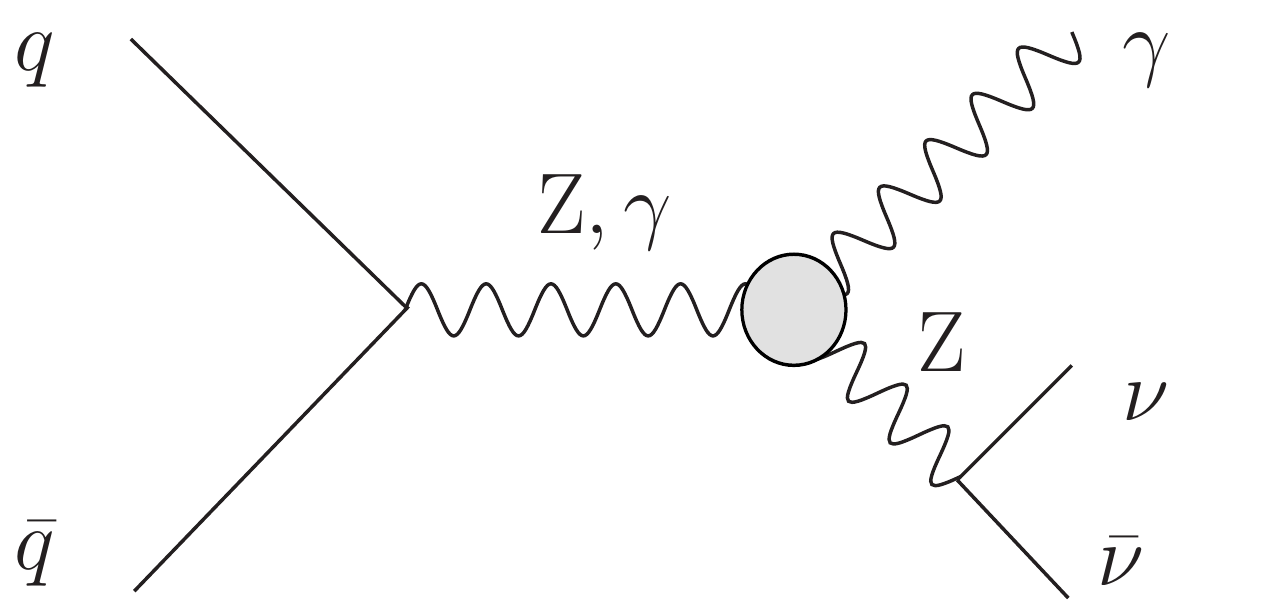}
\caption{Feynman diagrams of $\cPZ\gamma$ production via initial-state radiation in the SM at tree level (\cmsLeft), and via anomalous
$\cPZ\cPZ\gamma$ or $\cPZ\gamma\gamma$ trilinear gauge couplings
(\cmsRight).}
\label{fig:feynman_zg}
\end{center}
\end{figure}

The $\cPgn\cPagn\gamma$ final state can be produced through initial-state
radiation (where a photon is emitted by an initial-state parton) or through
anomalous coupling vertices.
The allowed electroweak tree-level diagram in the SM for $\cPZ\gamma$
production in pp collisions is shown in Fig.~\ref{fig:feynman_zg}~(\cmsLeft).
The $s$-channel production via a $\cPZ\cPZ\gamma$
or $\cPZ\gamma\gamma$ aTGC is shown in Fig.~\ref{fig:feynman_zg}~(\cmsRight).

The most general Lorentz-invariant and gauge-invariant
$\cPZ\mathrm{V}\gamma$ vertex can be described by four coupling parameters
$h_{i}^\mathrm{V}$ $(i = 1, \ldots, 4)$~\cite{baur, baur1}. The first two couplings ($i = 1, 2$) are
CP-violating, while the latter two ($i = 3, 4$) are CP-conserving~\cite{baur1, baur1993}.
At tree level in the SM, the individual values of these aTGCs are zero.
The photon transverse momentum spectrum has similar sensitivity
to CP-violating and CP-conserving couplings.
The results are generally interpreted in terms of the CP-conserving
aTGCs $h_3^\mathrm{V}$ and $h_4^\mathrm{V}$.

The sensitivity to aTGCs in $\cPZ\gamma$ production
is higher in the $\cPZ\to\cPgn\cPagn$
decay mode than in $\PZ$ boson decay modes with charged leptons,
because the branching fraction for a $\PZ$ boson decay to a pair of neutrinos
is six times higher than for a decay to a particular charged lepton pair,
and the acceptance in the neutrino channel is higher.

The fiducial phase space for this measurement is defined by the requirements of photon transverse energy $E_\mathrm{T}^{\gamma}>$ 145\GeV and photon pseudorapidity $\abs{\eta^{\Pgg}} <1.44$, where the contamination from other particles misidentified as photons is lower~\cite{Photon_perf}.

\section{The CMS detector}

The central feature of the CMS apparatus is a superconducting solenoid of 6\unit{m} internal diameter, providing a magnetic field of 3.8\unit{T}. Within the superconducting solenoid volume are a silicon pixel and strip tracker, a lead tungstate crystal electromagnetic calorimeter (ECAL), and a brass and scintillator hadron calorimeter (HCAL), each composed of a barrel ($\abs{ \eta }< 1.479$) and two endcap ($1.479 <\abs{ \eta } < 3.0$) sections, where $\eta$ is the pseudorapidity. Extensive forward calorimetry complements the coverage provided by the barrel and endcap detectors. Muons are measured in gas-ionization detectors embedded in the steel flux-return yoke outside the solenoid. The energy resolution for photons with
transverse momentum ${\ge}$ 60\GeV varies between 1\% and 2.5\% over the solid angle of the ECAL barrel, and from 2.5\% to 3.5\% in the endcaps~\cite{Photon_perf}. The timing measurement of the ECAL has a resolution better than 200\unit{ps} for energy deposits larger than 10\GeV~\cite{Photon_perf}. In the $\eta$-$\phi$ plane, where $\phi$ is the azimuthal angle and for $\abs{\eta}< 1.48$, the HCAL cells map onto $5 \times 5$ arrays of ECAL crystals to form calorimeter towers projecting radially outward from the nominal interaction point.

The event reconstruction is performed using a particle-flow (PF) algorithm~\cite{CMS-PAS-PFT-09-001,CMS-PAS-PFT-10-002}, which reconstructs and identifies individual particles using an optimized combination of information from all subdetectors. Photons are identified as energy clusters in the ECAL. These energy clusters are merged to form superclusters that are five crystals wide in $\eta$, centered around the most energetic crystal, and have a variable width in $\phi$. The energy of charged hadrons is determined from a combination of the track momentum and the corresponding ECAL and HCAL energies, corrected for the combined response function of the calorimeters. The energy of neutral hadrons is obtained from the corresponding corrected ECAL and HCAL energies. For each event, hadronic jets are formed from these reconstructed particles with the infrared- and collinear-safe
anti-$\kt$ algorithm~\cite{Cacciari:2008gp}, using a distance parameter $\Delta R = 0.5$, where $\Delta R = \sqrt{\smash[b]{(\Delta\eta)^{2} + (\Delta\phi)^{2}}}$
and $\Delta\eta$ and $\Delta\phi$ are the pseudorapidity and azimuthal angle difference between the jet axis and the particle direction.
The missing transverse momentum vector $\vec{\Em}_{\mathrm{T}}$ is defined as the projection on the plane perpendicular to the beams of the negative vector sum of the momenta of all reconstructed PF candidates in an event; its magnitude is referred to as \ETslash.

A more detailed description of the CMS detector, together with a definition of the coordinate system used and the relevant kinematic variables, can be found in Ref.~\cite{JINST1}.

\section{Signal and background modeling}

The final state consisting of an energetic photon accompanied by an imbalance in transverse energy can be mimicked by several other processes in the SM.
These processes include $\PW\Pgg\to\ell\Pgn\Pgg$ where $\ell$ is a charged lepton (if the lepton escapes detection), $\PW\to\ell\Pgn$ (if the lepton is misidentified as a photon), $\Pgg$+jets (if the jets are misreconstructed, resulting in \ETslash), QCD multijet production including $\cPZ(\nu\PAGn)$+jets (if the jet is misidentified as a photon), $\PZ\Pgg\to\ell\ell\Pgg$ (if both leptons escape detection), $\gamma\gamma$ events (if one of the photons escapes detection), and also backgrounds from beam halo.

The contributions from the $\PW\Pgg\to\ell\Pgn\Pgg$, $\Pgg$+jet, $\PZ\Pgg\to\ell\ell\Pgg$, and $\gamma\gamma$ processes to the candidate event sample are estimated using Monte Carlo-based (MC) simulations. The $\PW(\ell\nu)\gamma$ and $\PZ\to \ell\ell\gamma$ samples
are generated with \MADGRAPH{}5v1.3.30 at leading order (LO)~\cite{Madgraph} and then
processed with the \PYTHIA6.426 event generator~\cite{Pythia6} for showering and hadronization.
The other samples are generated with the \PYTHIA6.426 generator~\cite{Pythia6} at LO. All the samples are generated using the CTEQ6L1~\cite{CTEQ6L1} parton distribution function (PDF) set, processed through the CMS detector simulation based on \GEANTfour~\cite{GEANT, GEANTdev}, and reconstructed in the same manner as collision data.

The cross section for the SM background process $\PW\Pgg\to\ell\Pgn\Pgg$ with at most one jet is corrected with an
$E_\mathrm{T}^{\gamma}$ dependent K factor estimated from \MCFM~\cite{MCFM} to
account for next-to-leading-order (NLO) effects. The PDF4LHC Working Group recommendations~\cite{Alekhin:2011sk,PDF4LHC,PDF4LHC1} are used to estimate the uncertainty in the central value
of the NLO cross section arising from the PDFs, the strong coupling constant $\alpha_{s}$, and its scale dependence.
The $\Pgg$+jet cross section is corrected to include NLO effects.

To determine the efficiency for the SM $\cPZ(\nu\PAGn)\gamma$ production cross section measurement, events are produced with the \MADGRAPH{}5v1.3.30 generator at LO with a maximum of two additional partons and simulated through the full reconstruction chain.
Simulated samples of the $\cPZ\gamma$ signal for a grid of aTGC values are produced
using the \SHERPA v1.2.2 generator~\cite{sherpa}.
The cross section with at most one extra parton is corrected with an
$E_\mathrm{T}^{\gamma}$ dependent K factor estimated from \MCFM~\cite{MCFM} to
account for NLO effects.
The inclusive measurement has been compared with a theoretical calculation accurate up to next-to-next-to-leading order (NNLO).

To account for differences arising from imperfect modeling of the data in
the simulation, a total correction factor $\rho$ of $0.94\pm0.06$ is applied to all
MC-based background estimates. This is the product of individual correction factors defined as
ratios of the efficiencies measured in data and in simulation. They include $0.97\pm0.02$ for photon identification measured using $\cPZ\to\Pe\Pe$ events,
$0.99\pm0.03$ for timing requirements measured using a sample of electron
events, and $0.99\pm0.02$ and $0.99\pm$ 0.05 for lepton and jet vetoes measured using $\PW\to\Pe\Pgn$ events.

\section{Event selection}
Events are selected using both a single-photon trigger that requires a photon with $E_\mathrm{T}^{\gamma}> 150\GeV$,
and photon+\ETslash triggers with $E_\mathrm{T}^{\gamma}> 70\GeV$ and $\ETslash > 100\GeV$.
The combination of these triggers is
96\% efficient for events with photon transverse energy $E_\mathrm{T}^{\gamma}> 145\GeV$, photon pseudorapidity $\abs{\eta^{\Pgg}} <1.44$, and $\ETslash > 140\GeV$.
Events are required to have at least one primary vertex reconstructed within a longitudinal distance of $\abs{z} < 24\unit{cm}$  of the center of the
detector and at a distance ${<}2\unit{cm}$ from the $z$ axis. The primary vertex is chosen to be the vertex with the highest $\pt^{2}$ sum of its
associated tracks, where $\pt$ is the transverse momentum.

We impose additional requirements on the energy deposits in the calorimeters to distinguish photons from misidentified jets~\cite{Photon_perf}.
The energy in the HCAL associated with the photon supercluster should not exceed 5\% of its energy as measured in the ECAL. Moreover, the photon
candidates must have a shower distribution in the ECAL consistent with that expected for an
electromagnetic (EM) shower~\cite{Photon_perf}. To further reduce photon contamination arising
from misidentified jets, isolation requirements on photon candidates are imposed. Energy deposits for isolation
are obtained by considering particles in a cone around the axis defined by the
supercluster position and the primary vertex~\cite{Photon_perf}. In particular, the scalar sum
of transverse momenta (in \GeVns{}) of all photons within a cone of $\Delta R = 0.3$ around the supercluster, excluding a strip of width in $\eta$ of 0.015, is required to be less than $0.7+0.005 p_\mathrm{T}^{\gamma}$; the scalar sum of the transverse momenta (in \GeVns{}) of all charged hadrons, associated with the primary vertex, within a hollow cone of $0.02 < \Delta R < 0.30$ around the supercluster is
required to be less than $1.5$; and the scalar sum of the transverse momenta (in \GeVns{}) of all neutral hadrons within a cone
of $\Delta R = 0.3$ around the supercluster is required to be less than $1.0+0.04 p_\mathrm{T}^{\gamma}$.
Due to the large number of additional proton-proton interactions (pileup) in the same bunch crossing at the LHC, it is difficult
to know the true origin of the photon for a $\gamma$+\ETslash final state (our estimate is correct 50\% of the time), which could lead to an underestimation of isolation values.
Therefore, an additional PF-based charged particle isolation is calculated for each vertex and the largest
value of this isolation sum is required to be smaller than the nominal threshold used for charged particle isolation.

Photon candidates are required to have the energy deposited in the highest energy crystal
within the EM cluster to be within $\pm$3\unit{ns} of the time expected
for particles from a collision. This requirement reduces instrumental background arising from showers induced by bremsstrahlung from muons in the beam halo or in cosmic rays.
To further reduce this background, we exploit the characteristic
signature of showers from beam halo in the ECAL. A search region is defined around the highest energy crystal of
the EM cluster in a narrow $\phi$ window and over a wide $\eta$ range, after removal of the EM shower in a $5{\times}5$ array.
A straight line, parallel to the beam direction, is fitted over the remaining cells within this region. Events are
tagged as minimum ionizing particle (MIP tag) if the total energy deposited in the crystals associated with the straight-line fit is greater than
6.3\GeV.

Spurious signals can be embedded within EM showers by direct ionization of the avalanche
photodiode sensitive volume by highly ionizing particles. These signals, which would otherwise pass the EM shower selection criteria, are eliminated by requiring consistency among the energy deposition times for all crystals within an
EM shower.

Photon candidates are also removed if they are likely to be electrons, as inferred from patterns of hits in the pixel detector, called ``pixel seeds'', that are matched to the EM clusters~\cite{Electron_2015}.

Events containing good photon candidates are then required to have $\ETslash > 140\GeV$. A topological requirement of
$\Delta \phi > 2$ rad between the direction of the photon candidate and the vector $\vec{\Em}_{\mathrm{T}}$ is applied to reduce the contribution from the $\Pgg$+jet background.

In order to suppress backgrounds from QCD multijet production and leptonic decay of $\PW/\Z$+jets, events are vetoed if they contain significant hadronic/leptonic activity defined by:
(i) more than one jet with $\pt > 30\GeV$ not passing the pileup jet identification criteria~\cite{CMS-PAS-JME-13-005},
separated from the photon by $\Delta R > 0.5$, or (ii) an electron or a muon with
$\pt > 10\GeV$ and separated from the photon by $\Delta R > 0.5$.

To reduce the contamination from events with \ETslash arising from instrumental effects,
a $\chi^{2}$ function is constructed and minimized
\begin{equation}\label{eq:1a}
   \chi^2 = \sum_{i = \text{photon, jets}} \left( \frac{(\PT^{\text{reco}})_i-(\widetilde{p}_{\mathrm T})_i}{(\sigma_{\PT})_i} \right)^2 + \left( \frac{\widetilde{\Em}_x}{\sigma_{\widetilde{\Emsub}_x}}\right)^2  + \left( \frac{\widetilde{\Em}_y}{\sigma_{\widetilde{\Emsub}_y}}\right)^2,
\end{equation}
where the sum runs over the photon and all the jets in the event. The $(\sigma_{\PT})_{i}$ are the expected momentum resolutions of the reconstructed (reco) photon and jets, and the $(\widetilde{p}_{\mathrm T})_{i}$ are the free parameters allowed to vary in order to minimize the function. The resolution parametrization associated with the \ETslash is obtained from Ref.~\cite{metreso1}. Lastly, $\widetilde{\Em}_x$ and $\widetilde{\Em}_y$ are defined as
\begin{equation}
\begin{split}
\widetilde{\Em}_{x,y} &=  \Em^{\text{reco}}_{x,y} + \sum_{i = \text{photon, jets}} (p_{x,y}^{\text{reco}})_i - (\widetilde{p}_{x,y})_i
\ifthenelse{\boolean{cms@external}}{\\ &=}{=}
- \sum_{i = \text{photon, jets}} (\widetilde{p}_{x,y})_i, \\
\\
\widetilde{\Em}_{\mathrm{T}} &= \sqrt{\widetilde{\Em}_{x}^2 + \widetilde{\Em}_{y}^2}.
\end{split}
\end{equation}

For events with no true \ETslash, the $\chi^{2}$ is expected to be small, with values of $\widetilde{\Em}_{\mathrm{T}}$ close to 0, while for events with significant true \ETslash the minimization will result in high $\chi^{2}$ values, with $\widetilde{\Em}_{\mathrm{T}}$ close to the actual \ETslash in the event. An additional requirement of $\widetilde{\Em}_{\mathrm{T}} > 120$\GeV reduces the number of $\Pgg$+jet (QCD multijet) events by 80\% (35\%), while keeping 99.5\% of signal events.

After applying these requirements, 630 candidate events are observed in data.

\section{Background estimation}

The largest contribution is found in the $\PW\Pgg\to\ell\Pgn\Pgg$ process and is estimated to be
$103\pm21$ events. The contributions from other processes, a small fraction of the total background, amount to $36\pm3$ events.

The most significant background contribution estimated using simulation
is also validated in a control region dominated by $\PW(\ell\nu)\gamma$ events.
Events are selected using the full candidate selection but with the
lepton veto inverted. In data, 104 events are observed, consistent with an
expectation of $126\pm23$ events.

The background originating from jets misidentified as photons is estimated using a data driven method. The method is based on a class of jets, referred to as ``photon-like'' jets, that have properties similar to electromagnetic objects. Photon-like jets are required to pass a very loose photon selection but at the same time fail one of the isolation requirements. The method also relies on the ratio of jets passing the full photon selection to those identified as photon-like jets. This ratio is measured in a control sample enriched in QCD multijet events. To suppress the contribution of electroweak processes, the missing transverse energy in this control sample is required to be smaller than 30\GeV. Because this sample also contains true isolated photons from QCD direct photon production, this contribution must be subtracted from the numerator of the ratio. The required correction is estimated by performing a fit to the distribution of the candidate shower width variable $\sigma_{\eta \eta}$~\cite{Photon_perf}. Two shower shape profiles are used in this fit, the shower shape of true photons, obtained from simulated $\Pgg$+jet events, and the shower shape of photon-like jets, obtained from the charged hadron isolation sideband in data. This corrected ratio is used to weight a set of data events where the photon candidate passes the photon-like jet selection criteria. The estimated number of background events is found to be $45\pm14$, where the uncertainty reflects an uncertainty in the estimation of the ratio, as well as the statistical uncertainty of the sample scaled for the final estimate.

An instrumental background caused by electrons arises due to the imperfect efficiency
for reconstructing and associating pixel seeds with
clusters. For our kinematic requirements, this background largely originates
from $\PW$ boson ($\PW \to \Pe\nu$) production, and is estimated from data.
The pixel seed efficiency $\epsilon_\text{pix}$ is measured in $\PZ\to\Pe\Pe$ events using the
standard ``tag-and-probe'' method~\cite{tp} and is estimated to
be $0.984\pm0.002$ for electrons with $E_\mathrm{T} > 100\GeV$.  To
estimate the final yield of this background, a factor of
(1-$\epsilon_\text{pix}$)/$\epsilon_\text{pix}$ is applied to a set of events
in the data with the same candidate event selection as the signal candidates and
with the additional requirement of a pixel seed match. The resulting contribution is
estimated to be $60\pm6$ events, where the uncertainty
is dominated by the uncertainty in the measurement of
$\epsilon_\text{pix}$.

Since photon candidates are only identified within the ECAL,
the candidate sample is susceptible to contamination from
noncollision backgrounds. These backgrounds arise from interactions in the calorimeter
of accelerator related particles (beam halo), spurious signals
in the ECAL itself, and particles originating from cosmic ray
interactions. The timing distribution measured from the
ECAL for each of these backgrounds is distinctly different from the arrival
time distribution for photons produced in collisions. A fit is performed
to the candidate time distributions using shapes derived from data.
The background distribution are constructed by inverting MIP tag (beam halo)
and shower shape (anomalous signal) requirements. The arrival time for
photons from the interaction region is modeled using $\PW\to\Pe\nu$
candidates in data. From the result of the fit, the only
significant noncollision background is found to be from beam halo events,
and its contribution is estimated to be $25\pm6$ events.

The total number of expected background events is
$269 \pm 26$, as mentioned in Table~\ref{tab:BkgSummaryC}.
The number of signal events (data - expected background) is $361 \pm 36$, where the uncertainty is obtained by adding in quadrature the uncertainty from the
data and the background estimation.
The expected number
of $\PZ\Pgg\to\Pgn\Pagn\Pgg$ signal events,
obtained using \MADGRAPH{}5 and corrected for NNLO effects, is $345 \pm 43$.

\section{Cross section measurement}

A summary of the backgrounds and data yields is given in
Table~\ref{tab:BkgSummaryC}, wherein the uncertainties
in the background estimates include
both statistical and systematic sources.

\begin{table}[htb]
\center
\topcaption{Summary of estimated $\PZ(\to\nu\PAGn)+\gamma$ signal, backgrounds, and observed total number of candidates. Backgrounds listed as ``Others'' include the small contributions from $\PW\to \mu\Pgn$, $\PZ\Pgg\to \ell\ell\Pgg$, $\gamma\gamma$, and $\gamma$+jet. Uncertainties include both statistical and systematic contributions.
\label{tab:BkgSummaryC}}
{
\newcolumntype{x}{D{,}{\,\pm\,}{-1}}
\begin{tabular}{l|x}
\hline
\multicolumn{1}{l|}{Process} & \multicolumn{1}{c}{Estimate} \\
\hline
$\PW(\to \ell\nu)+\gamma     $ & 103 , 21   \\
$\PW\to \Pe\nu $ & 60  , 6   \\
${\rm jet}\to\gamma~{\rm MisID}$ & 45 , 14 \\
Beam halo &  25 ,  6 \\
Others & 36 , 3  \\
\hline
Total background                       &  269 , 26 \\
\hline
$\PZ(\to\nu\PAGn)+\gamma    $ & 345 , 43  \\
\hline
Data                                   & \multicolumn{1}{c}{630}    \\
\hline
Data - background    & 361 , 36 \\
\hline
\end{tabular}
}
\end{table}

The $\cPZ\gamma \to \cPgn\cPagn\gamma$ cross section for
$E_\mathrm{T}^{\gamma}> 145\GeV$ and $\abs{\eta}^{\gamma}  < 1.44$ is calculated
using the following formulae:
$$ \sigma  \mathcal{B}= \frac{N_\text{data}-N_\text{bkg}}{A \epsilon \, L},$$
$$A \epsilon = ({A\epsilon})_\text{sim} \, \rho,$$
where $N_\text{data}$ is the number of observed events, $N_\text{bkg}$
is the estimated number of background events, $A$ is the geometrical acceptance, $\epsilon$ is the selection efficiency to select inclusive $\PZ(\to\nu\PAGn)+\gamma$ events offline, and $L$ is the integrated
luminosity. The product of ${A\epsilon}$ is
estimated from the simulation to be $0.377 \pm 0.001$, where the
uncertainty is statistical. $\rho$ is the correction factor defined in Section 3.

The photon, jet and \ETslash energy scales and resolutions, pileup, correction factor $\rho$, and the uncertainties in the PDFs are considered as sources of systematic uncertainty in the acceptance calculation.
The uncertainty in the photon energy scale is about $1.5\%$, which translates into an uncertainty in $A \epsilon$ of $^{+3.4}_{-5.0}$\%, where $A$ is the geometrical and
kinematic acceptance of the selection criteria, and $\epsilon$ is the
signal selection efficiency.
Additionally, there are systematic uncertainties due to the jet energy scale and jet resolution in the measurement of \ETslash, which give $^{+2.3}_{-2.3}$\%
and $^{+1.2}_{-1.4}$\%, respectively, and the unclustered energy scale, which gives $^{+1.9}_{-0.6}$\%.
For pileup, a central value for the total inelastic cross section of 69.4 mb~\cite{puxsec1,puxsec2} is used.
A variation of ${\pm}5\%$ in the number of interactions
is used to cover the uncertainty in $A\epsilon$ due to pileup modeling, which is 0.3$\%$. The uncertainty in the integrated luminosity~\cite{CMS-PAS-LUM-13-001} is 2.6\%.
Other sources include the uncertainty in the correction factor $\rho$, which contributes 6.4\%.

A summary of the systematic uncertainties in $A \epsilon$ for the $\PZ(\nu\PAGn)\gamma$ signal sample is shown in Table~\ref{tab:sysfull}.

\begin{table}[tbp]
\centering
\topcaption{Systematic uncertainties considered in $A \epsilon$ for the $\PZ(\nu\PAGn)\gamma$ signal sample from various sources. }
\small
\begin{tabular}{l|c}
\hline
\multicolumn{1}{l|}{Source}    &  $\PZ(\nu\PAGn)\gamma$ [$\%$]  \\
\hline
Photon and \ETslash energy scale  & +3.4, -5.0  \\
Jet and \ETslash energy scale   & $\pm$2.3\\
Jet energy resolution   & $\pm$1.3\\
Unclustered energy   & $\pm$1.2\\
Pileup        & $\pm$0.3\\
Luminosity &          $\pm$2.6\\
Correction factor $\rho$ & $\pm$6.4\\
\hline
\end{tabular}
\label{tab:sysfull}
\end{table}

The measured production cross section $\sigma(\Pp\Pp\to \PZ\Pgg) \, \mathcal{B} (\PZ \to\Pgn\Pagn)$
for
$E_\mathrm{T}^{\gamma}>$ 145\GeV and $\abs{\eta} < 1.44$
is $52.7 \pm 2.1\stat \pm 6.4\syst \pm 1.4\lum\unit{fb}$.

The expected cross section of the signal process for
$E_\mathrm{T}^{\gamma}> 145\GeV$ and $\abs{\eta}^{\gamma} < 1.44$, obtained with
the NLO generator \MCFM, is $40.7\pm4.9 \unit{fb}$.
The quoted uncertainty in the
prediction takes into account the PDF and scale uncertainties.
The NNLO theoretical prediction~\cite{Grazzini,Grazzini1} is $50.0^{+ 2.4}_{- 2.2}\unit{fb}$,
where the uncertainty includes only scale variations.

\begin{figure*}[htb]
\begin{center}
\includegraphics[width=0.45\textwidth]{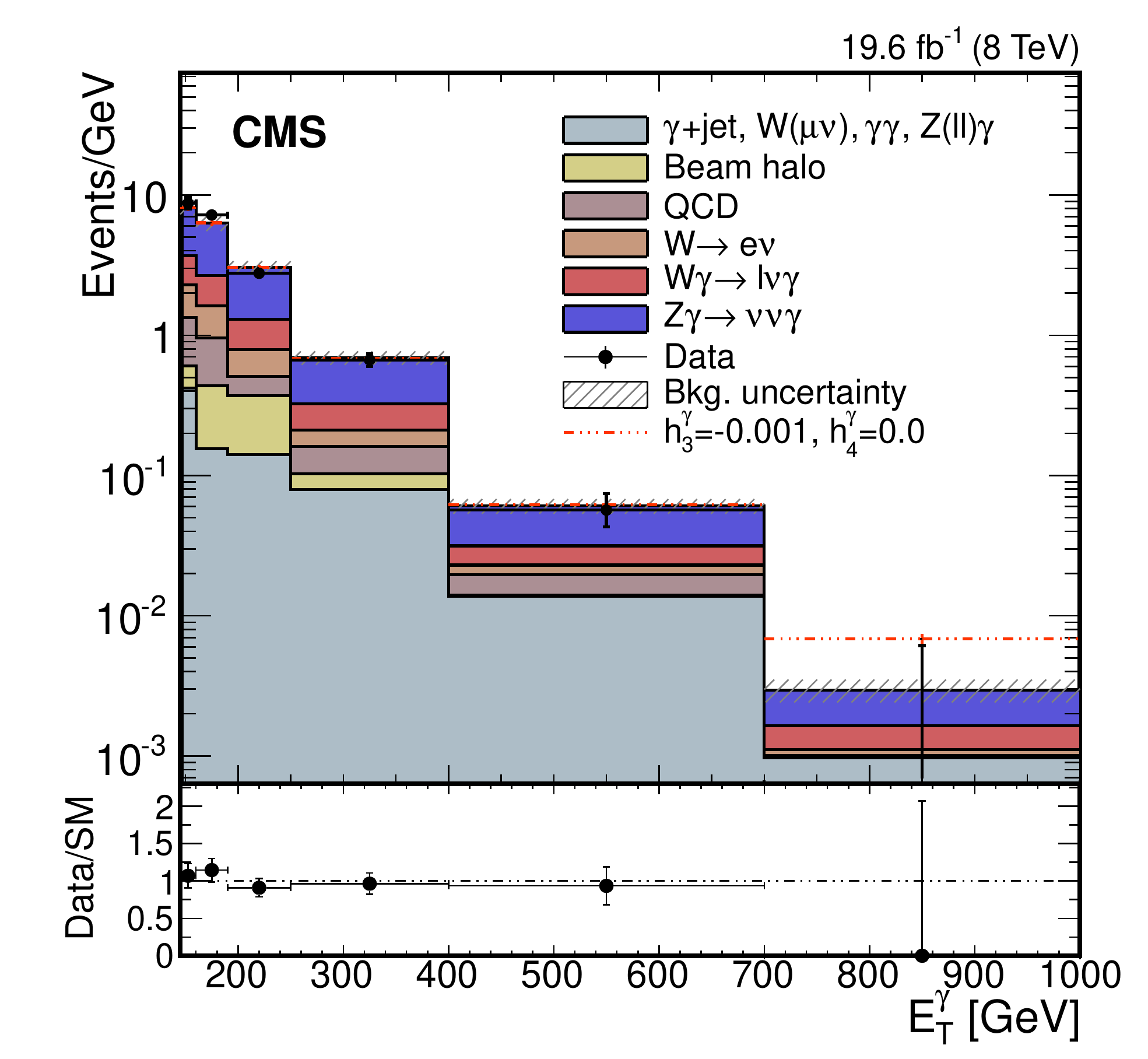}
\includegraphics[width=0.45\textwidth]{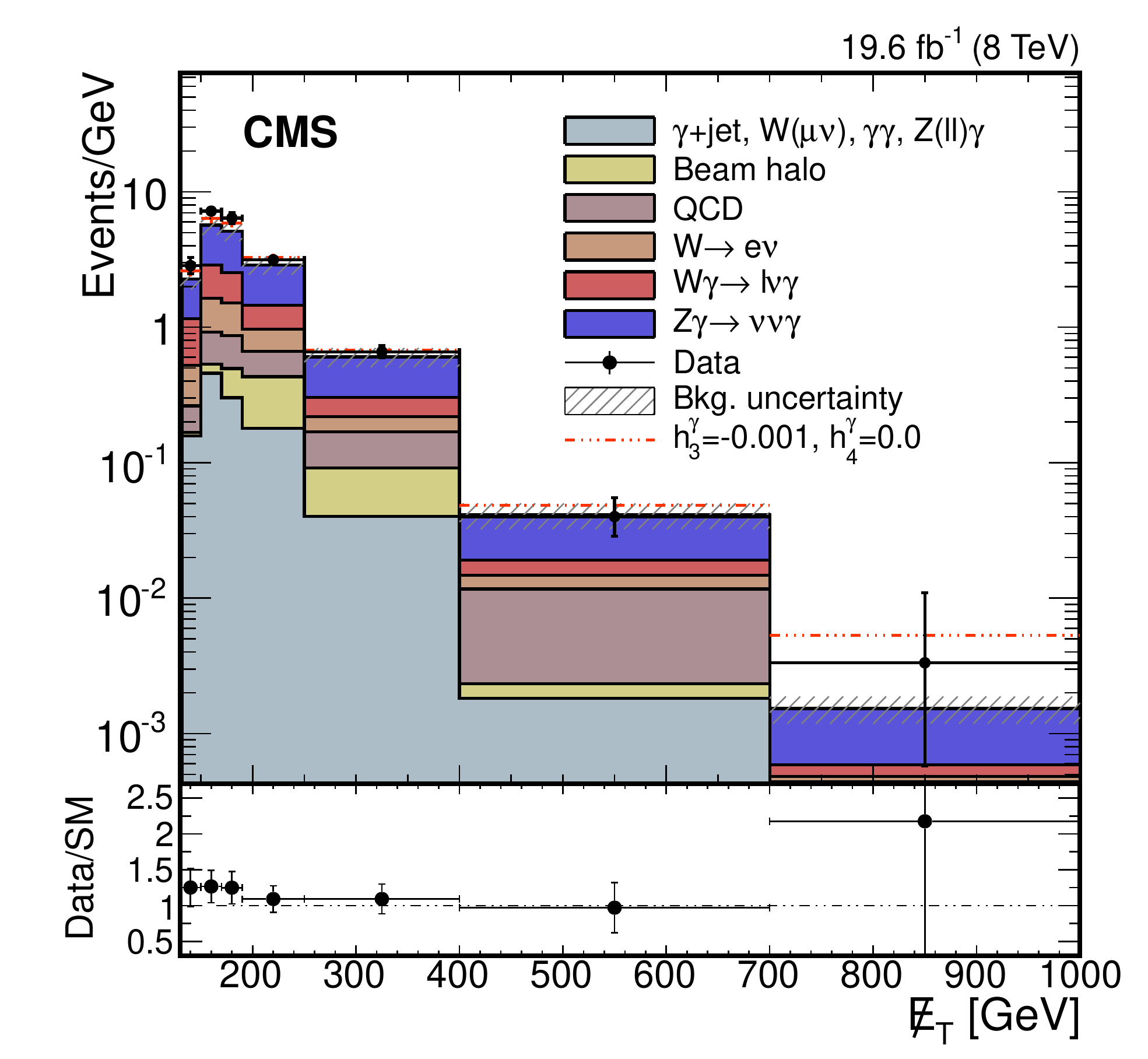}
\caption{The $E_\mathrm{T}^{\gamma}$ and \ETslash distributions in data
  (points with error bars) compared with the SM
  $\PZ\Pgg\to\Pgn\Pagn\Pgg$ signal and
  estimated contributions from backgrounds.
  A typical aTGC signal from $\cPZ\gamma\gamma$ with $h_{3}^\gamma$ = -0.001, $h_{4}^\gamma$ = 0.0  would provide an excess, as shown in the
  dot-dashed histogram. The background uncertainty includes statistical and systematic components.}
  \label{fig:photon_pt}
\end{center}
\end{figure*}

The distributions of photon transverse energy and \ETslash are shown in
Fig.~\ref{fig:photon_pt}, with the signal and background predictions overlaid.
The expected contribution from a $\cPZ\gamma\gamma$ aTGC signal with $h_{3}^\gamma$ = -0.001, $h_{4}^\gamma$ = 0.0 is also shown.
No significant excess of events over the SM expectation is observed.

\section{Limits on trilinear gauge couplings}
We use the $\ET^\gamma$ spectrum to set limits on aTGCs
by means of a likelihood formalism. In this study, we follow the CMS convention of not
suppressing the aTGCs by an energy-dependent form factor.

The probability of observing the number of data events in a
given range of $\ET^\gamma$ is estimated using a
Poisson distribution given by the expected
signal and background predictions.
Limits on aTGCs are calculated on the
basis of a profile likelihood method as described in
Ref.~\cite{prlik}.
In the fit to the observed spectra, systematic uncertainties are represented
 by nuisance parameters with log-normal prior probability density functions. The changes in shape of the observed spectra that result from varying the photon energy scale and the
theoretical differential cross section within their respective uncertainties are treated using a morphing technique~\cite{vmor}.

\begin{figure}[htb]
\begin{center}
{\includegraphics[width=0.45\textwidth]{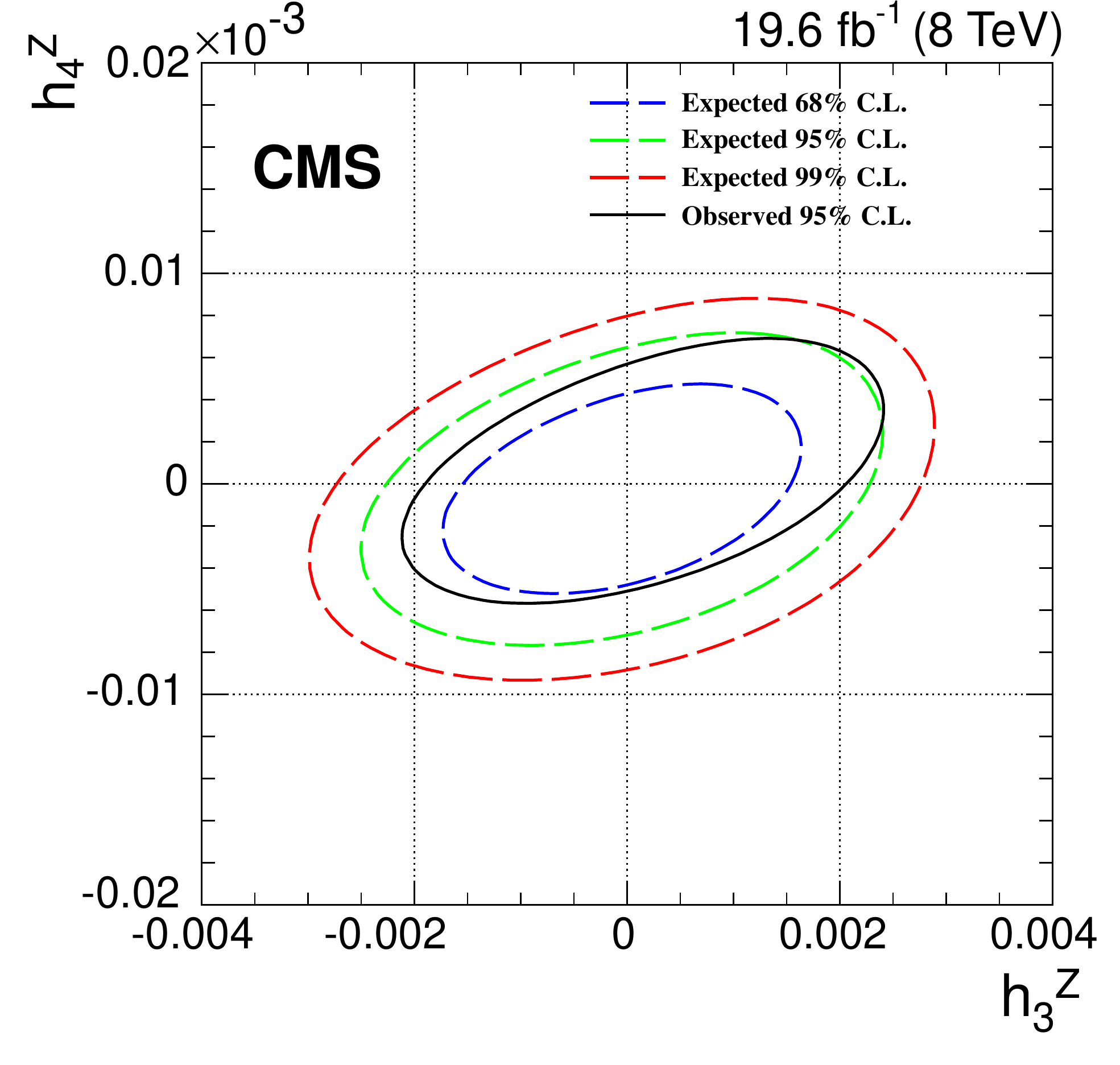}}
\caption{Two-dimensional 95\% CL limits on $\cPZ\cPZ\gamma$ couplings.}
\label{fig:2dlimits}
\end{center}
\end{figure}

\begin{figure}[htb]
\begin{center}
{\includegraphics[width=0.45\textwidth]{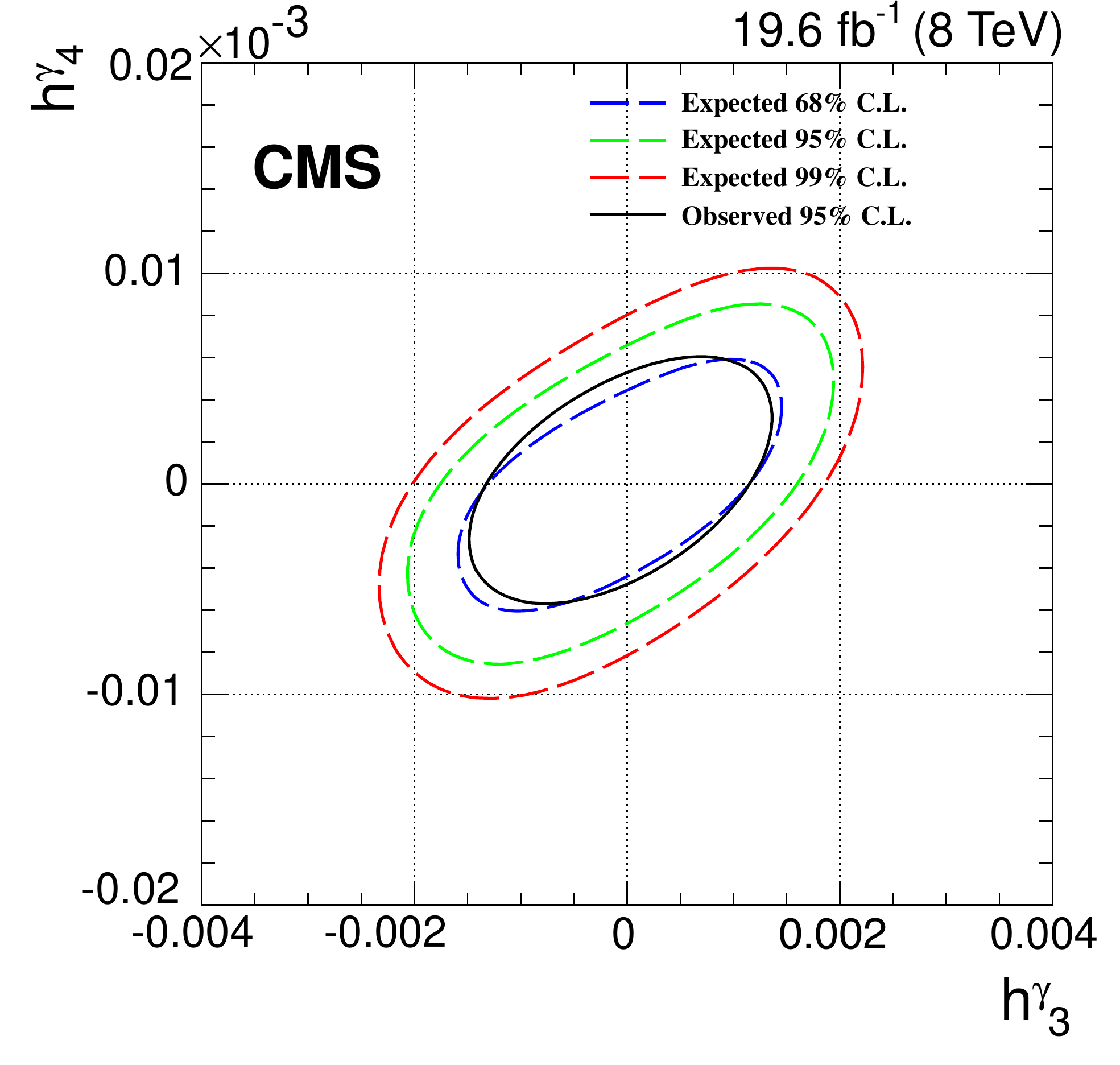}}
\caption{Two-dimensional 95\% CL limits on $\cPZ\gamma\gamma$ couplings.}
\label{fig:2dlimits1}
\end{center}
\end{figure}

The best fit value from data for the aTGCs is very close to the SM values.

Limits at 95\% confidence level (CL) are set on pairs of aTGC parameters $(h_{3}^{\cPZ}$,
$h_{4}^{\cPZ})$ and $(h_{3}^\gamma$, $h_{4}^\gamma)$, as presented in
Fig.~\ref{fig:2dlimits} and Fig.~\ref{fig:2dlimits1}, respectively.
Furthermore, one-dimensional 95$\%$ CL limits
are obtained for a given aTGC while setting the other neutral aTGCs
to their SM values, \ie, to zero.
A summary of the one-dimensional limits along with 7\TeV is given in Table \ref {tab:atgcLimitsN_0}.

\begin{table*}[htb]
  \begin{center}
   \topcaption{One-dimensional 95\% CL limits on $\PZ\mathrm{V}\gamma$ anomalous trilinear gauge couplings from the $\PZ\Pgg\to\Pgn\Pagn\Pgg$ channel. The limits obtained from data with $\sqrt{s} = 7$ \TeV are also shown.}
    \begin{tabular}{c|c|c}
      \hline
      Coupling           & $\sqrt{s} = 8\TeV$   & $\sqrt{s} = 7\TeV$  \\
      \hline
      $h_3^{\cPZ}$        & [-1.5, 1.6]$\times 10^{-3}$   & [-2.7, 2.7]$\times 10^{-3}$         \\
      $h_4^{\cPZ}$     & [-3.9, 4.5]$\times 10^{-6}$    & [-1.3, 1.3]$\times 10^{-5}$     \\
      $h_3^\gamma$   & [-1.1, 0.9]$\times 10^{-3}$ & [-2.9, 2.9]$\times 10^{-3}$  \\
      $h_4^\gamma$   & [-3.8, 4.3]$\times 10^{-6}$ & [-1.5, 1.5]$\times 10^{-5}$  \\
      \hline
    \end{tabular}
    \label{tab:atgcLimitsN_0}
  \end{center}
\end{table*}

\section{Summary}
We have presented an inclusive measurement of the
$\PZ\Pgg\to\Pgn\Pagn\Pgg$ production cross section in pp collisions
at $\sqrt{s} = 8\TeV$ using data collected with the CMS experiment in 2012, corresponding to an integrated luminosity of 19.6\fbinv.
The measured cross section $\sigma(\Pp\Pp\to \PZ\Pgg) \, \mathcal{B} (\PZ \to\Pgn\Pagn)$ for
photons with $E_\mathrm{T}^{\gamma}>$ 145\GeV and $\abs{\eta^{\gamma}} <1.44$ is
$52.7 \pm 2.1\stat\pm 6.4\syst\pm 1.4\lum\unit{fb}$,
in agreement with the NNLO prediction~\cite{Grazzini,Grazzini1} of
$50.0^{+ 2.4}_{- 2.2}\unit{fb}$.
No evidence was found for anomalous neutral trilinear gauge couplings
in $\PZ\gamma$ production. Limits at 95\% CL were placed on the
$h_3^\mathrm{V}$ and $h_4^\mathrm{V}$
parameters of $\cPZ\cPZ\gamma$ and $\cPZ\gamma\gamma$ couplings:

\begin{center}
\linespread{2.0}
 $ -1.5 \times 10^{-3} < h_3^{\cPZ}  < 1.6 \times 10^{-3}$\\
\linespread{2.0}
 $ -3.9 \times 10^{-6} < h_4^{\cPZ}  < 4.5 \times 10^{-6}$\\
 $ -1.1 \times 10^{-3} < h_3^\gamma  < 0.9 \times 10^{-3}$\\
 $ -3.8 \times 10^{-6} < h_4^\gamma  < 4.3 \times 10^{-6}$.
\end{center}

These results yield the most stringent limits to date on anomalous neutral trilinear gauge couplings.

\section*{Acknowledgements}
We thank Massimiliano Grazzini and Dirk Rathlev for providing us with the NNLO calculation of the cross section.
We congratulate our colleagues in the CERN accelerator departments for the excellent performance of the LHC and thank the technical and administrative staffs at CERN and at other CMS institutes for their contributions to the success of the CMS effort. In addition, we gratefully acknowledge the computing centers and personnel of the Worldwide LHC Computing Grid for delivering so effectively the computing infrastructure essential to our analyses. Finally, we acknowledge the enduring support for the construction and operation of the LHC and the CMS detector provided by the following funding agencies: BMWFW and FWF (Austria); FNRS and FWO (Belgium); CNPq, CAPES, FAPERJ, and FAPESP (Brazil); MES (Bulgaria); CERN; CAS, MoST, and NSFC (China); COLCIENCIAS (Colombia); MSES and CSF (Croatia); RPF (Cyprus); MoER, ERC IUT and ERDF (Estonia); Academy of Finland, MEC, and HIP (Finland); CEA and CNRS/IN2P3 (France); BMBF, DFG, and HGF (Germany); GSRT (Greece); OTKA and NIH (Hungary); DAE and DST (India); IPM (Iran); SFI (Ireland); INFN (Italy); NRF and WCU (Republic of Korea); LAS (Lithuania); MOE and UM (Malaysia); CINVESTAV, CONACYT, SEP, and UASLP-FAI (Mexico); MBIE (New Zealand); PAEC (Pakistan); MSHE and NSC (Poland); FCT (Portugal); JINR (Dubna); MON, RosAtom, RAS and RFBR (Russia); MESTD (Serbia); SEIDI and CPAN (Spain); Swiss Funding Agencies (Switzerland); MST (Taipei); ThEPCenter, IPST, STAR and NSTDA (Thailand); TUBITAK and TAEK (Turkey); NASU and SFFR (Ukraine); STFC (United Kingdom); DOE and NSF (USA).

\bibliography{auto_generated}
\cleardoublepage \appendix\section{The CMS Collaboration \label{app:collab}}\begin{sloppypar}\hyphenpenalty=5000\widowpenalty=500\clubpenalty=5000\textbf{Yerevan Physics Institute,  Yerevan,  Armenia}\\*[0pt]
V.~Khachatryan, A.M.~Sirunyan, A.~Tumasyan
\vskip\cmsinstskip
\textbf{Institut f\"{u}r Hochenergiephysik der OeAW,  Wien,  Austria}\\*[0pt]
W.~Adam, E.~Asilar, T.~Bergauer, J.~Brandstetter, E.~Brondolin, M.~Dragicevic, J.~Er\"{o}, M.~Flechl, M.~Friedl, R.~Fr\"{u}hwirth\cmsAuthorMark{1}, V.M.~Ghete, C.~Hartl, N.~H\"{o}rmann, J.~Hrubec, M.~Jeitler\cmsAuthorMark{1}, V.~Kn\"{u}nz, A.~K\"{o}nig, M.~Krammer\cmsAuthorMark{1}, I.~Kr\"{a}tschmer, D.~Liko, T.~Matsushita, I.~Mikulec, D.~Rabady\cmsAuthorMark{2}, B.~Rahbaran, H.~Rohringer, J.~Schieck\cmsAuthorMark{1}, R.~Sch\"{o}fbeck, J.~Strauss, W.~Treberer-Treberspurg, W.~Waltenberger, C.-E.~Wulz\cmsAuthorMark{1}
% ...
\vskip\cmsinstskip
% ...
1:~~Also at Vienna University of Technology, Vienna, Austria\\
2:~~Also at CERN, European Organization for Nuclear Research, Geneva, Switzerland\\
% ...

\end{sloppypar}
\end{document}